# Deployment in dynamic environments

**Jose L. Ruiz\* — Juan C. Dueñas\* — Fernando Usero\*\* — Cristina Díaz\*\*\***

\* *Universidad Politécnica de Madrid*
*Madrid, Spain*
*{jlruiz, jcduenas}@dit.upm.es*

\*\* *Telvent Interactiva*
*Seville, Spain*
*fernando.usero@telvent.abengoa.com*

\*\*\* *Telefónica I+D*
*Madrid, Spain*
*cdd@tid.es*

*ABSTRACT: Information and communication technologies are moving towards a new stage where applications will be dynamically deployed, uninstalled, updated and (re)configured. Several approaches have been followed with the goal of creating a fully automated and context-aware deployment system. Ideally, this system should be capable of handling the dynamics of this new situation, without losing sight of other factors, such as performance, security, availability or scalability. We will take some of the technologies that follow the principles of Service Oriented Architectures, SOA, as a paradigm of dynamic environments. SOA promote the breaking down of applications into sets of loosely coupled elements, called services. Services can be dynamically bound, deployed, reconfigured, uninstalled and updated. First of all, we will try to offer a broad view on the specific deployment issues that arise in these environments. Later on, we will present our approach to the problem. One of the essential points that has to be tackled to develop an automated deployment engine will be to have enough information to carry out tasks without human intervention. In the article we will focus on the format and contents of deployment descriptors. Additionally, we will go into the details of the deployment framework for OSGi enabled gateways that has been developed by our research group. Finally we will give some concluding remarks and some ideas for future work.*

**1. Introduction**

Information and communication technologies are moving towards a new stage, where applications will be dynamically deployed, uninstalled, updated and (re)configured. Some of the main factors that are pushing and enabling this evolution are as follows:

- The consumer electronics industry is evolving quickly, following the pace of Moore's law - the computing power doubles every one and a half years -, therefore devices are capable of carrying out increasingly complex tasks by themselves.
- Economies of scale are cutting the cost of devices and communication services. Access to the Internet is ubiquitous and affordable.
- Services development cycles are becoming shorter and the product quality thus obtained improving. These are consequences of different facts: service developers have more powerful tools at hand (configuration management systems, modelling tools, integrated development environments and collaborative tools) and software engineering is reaching a quite mature state (design patterns, standardised procedures and so on).
- The Internet has broadened companies target market to a global scale. The ability to keep the products up with the market needs is essential to survive in this aggressive environment.

An effective deployment infrastructure is obviously a key issue in this context, as is the first step to offer services to the customers. Several approaches have been followed aiming at creating a fully automated and context aware deployment system. Ideally, this system should be capable of handling the dynamics of this new situation, without losing sight of other factors, such as performance, security, availability or scalability.

In this article we will discuss some of the technologies that follow the principles of Service Oriented Architectures, SOA[BIE 03][BOU 04], which is one of the main paradigms of dynamic environments and is therefore a good choice for obtaining generic conclusions. First of all, we will try to give an overview of the deployment issues that arise in these environments. In particular we will focus on the contents of deployment metadata files, which are essential to the work of nearly all deployment systems. After that, we will go into details on Resolvit, the deployment engine for OSGi[OSG 03] enabled gateways that has been developed by our research group. We will present Resolvit's deployment algorithm and its deployment descriptors. Finally we will give some concluding remarks and ideas for the future work.

**2. Deployment in Service Oriented Architectures**

SOA promote the breaking down of applications into sets of loosely coupled elements, called services. From a generic point of view a service may be understood as any piece of software that offers a certain functionality. A service must be defined by means of an interface. One or more implementations for the service may exist, but all of them will have to be compliant with the primitives declared in the service interface.

Service clients may be users or other services. One of the most characteristic features of SOA is the loose coupling between clients and services. In fact, the binding between them happens at run time, thanks to the presence of a service registry, see Figure 1. The client will look up a service, previously published in the registry, and then will access it. Both client and service may not be located at the same place, so communication protocols and connectivity will probably be needed. There are several technologies that follow the principles of SOA, among them the following can be mentioned: Web Services, OSGi and Jini. Each one has a different target domain; Jini is oriented to services in local area networks, Web Services care about interaction between services offered by different organizations through the Internet and OSGi is focused on the interaction between services deployed in local/personal networks (e.g. home domain), and entities located on the Internet. Our main concern is the OSGi platform and consequently special attention will be paid to it in the rest of the article.

In SOA, applications are dynamically created by interaction between services. This alludes to one of the main topics we want to address with this article; software dependencies. The term dependency is used by us in a broad sense to express any kind of relationship between software elements and not only in the sense of something that is needed to function. In SOA, dependencies are usually satisfied at runtime rather than at other stages: design, implementation and configuration. From a pure SOA point of view dependencies will be of two kinds:

- *Visible in the service interface*. Either as input or output parameters, the interface of one of the services will make use of the interface of another service or even of another service implementation.
- *Invisible in the service interface*. Some service implementations will rely on another services, either interfaces or implementations, to offer their functionality, but this fact is not made public in the interface.

Automated application deployment will have to deal with service dependencies resolution, both visible and invisible. Additional information will have to be provided to enable the work of an automated deployment engine, in order to tackle with invisible dependencies. This information can be obtained by automated introspection on the service implementations, but in most of the cases this process is costly in time and prone to errors. The alternative is to provide this information in metadata files[OMG 03][DEB 04a]. Following the principles of declarative programming, the metadata will contain the information about what is needed and offered. The deployment engine will be responsible for resolving this situation.

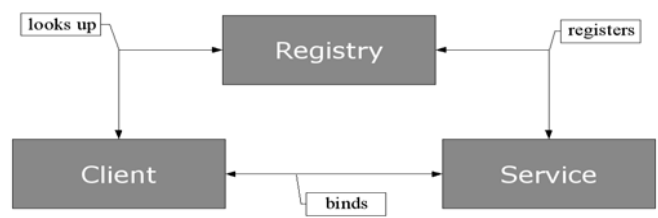

*Figure 1.* Service Oriented Architectures

The metadata files will be logically attached to service implementations. Physically they may be available in the same package and/or separately. Having the metadata available separately has the advantage that the dependencies analysis does not imply downloading the whole packages; this saves time and network traffic if these packages are not finally selected for deployment. Another advantage of this strategy is that this information can be cached in the deployment targets, in order to run the analysis phase locally, avoiding subsequent network calls. Debian's apt-get tool [DEB 04a] uses this approach. Service implementations will be made up of different kinds of resources. In the case of the OSGi platform, for example, service implementations are packaged in bundles, which may contain: Java classes, native libraries and other types of resources (text files, images and so on).

In order to successfully deploy applications in SOA, all dependencies, visible in the interfaces or invisible and consequently related to the implementations, will have to be solved. In the next section we look at the requirements that this imposes on the deployment information that will have to be considered.

## 3. Deployment requirements

In this section we are going to summarize the main requirements for deployment in service oriented architectures. We will focus on a concrete scenario: service deployment for OSGi enabled home services gateways. This scenario is rich enough to extract general requirements and it will give us the opportunity to illustrate the requirements with examples. Before going into details on the requirements themselves a brief summary of OSGi follows.

### *3.1. OSGi Open Services Gateway initiative*

Open Services Gateway Initiative (OSGi) is an industrial consortium launched in 1998 by several companies working in the embedded systems market. The initial objective of OSGi was to produce open specifications targeted to standardizing Java- based service platforms. Inside the OSGi consortium there is an interest group on home services gateways, for short service gateways. A service gateways acts as a bridge between local network/devices and wide area networks, i.e. the Internet. This is not the unique field of application of OSGi, though: this year, a group with the main focus on using the OSGI service platform in mobile devices has been founded and OSGi has been chosen as the core execution environment for Eclipse[ECL 04].

The OSGi service platform is made up of three layers, bottom up: a Java virtual machine, a basic execution environment called the OSGi framework and a set of standard services. As relevant features of the OSGi platform it is important to point out its support to handle different versions of services implementations at runtime, its adaptability (it can be deployed on platforms with different operating systems and limited in resources) and the inclusion of a service registry in the OSGi framework.

An OSGi service is defined by a Java interface and it will be implemented by a service object, which in turn may implement more than one service. In order to be used by others, the services will have to be already registered in the service registry. At registration time, a service implementation can include some specific information used to differentiate itself from other implementations. This information is defined by a set of key-value pairs, and can be modified at runtime by the service provider. Entities that desire to use a service can then choose between available implementations depending on these properties.

OSGi defines its own deployment unit for service implementations, known as a bundle. A bundle is a zip compressed file that contains: Java classes (service objects, implementations and utility classes), a bundle activator, one metadata file and other resources. The OSGi framework allows to manage the bundle life cycle, which is made up of the following states: installed, resolved, started, starting, stopping, stopped and uninstalled. The metadata file is interpreted by the OSGi framework, which validates if dependencies at the Java package level are resolved. A bundle can not be started if all its necessary classes are not available, these may be included in the same bundle or exported by other bundles. In case the bundle cannot be resolved nothing is done by the OSGi framework. No further dependency resolution is carried out.

One of the contributions of our work has been to provide the OSGi framework with the ability to find, verify, download and install those service implementations, Java libraries and/or native packages required for the deployment of any OSGi application.

### 3.2. Requirements

In order to promote resource saving, redundant deployments (installing the same elements more than once) must be avoided. This imposes a requirement on the deployment process: the *status of the platform must be remembered or stored* . Resolvit's architecture is designed to access the status of the platform either locally or remotely, if other server is responsible for its storage.

The SOA deployment process *must be complete and adaptive*. In order to deploy a service and make use of it, its implementation and other required resources (including other service implementations) will have to be deployed.

The requirement of completeness has two important consequences: the first one is that in order to automate dynamic application assembling we will need service discovery mechanisms. Means to specify the dependencies at service level are a must. As commented on the previous section, deployment descriptors are a good place to store this information. Deployment descriptors must be available separately, as this enables the analysis of these data to be carried out without downloading the whole deployment unit. Besides, having this information apart from the code promotes flexibility and role independence between the service developer and the application assembler. The second of the consequences is that the deployment process *must span its activity over multiple layers on the platform* and the deployment engine has to deal with dependencies at all of them. In the home services gateway scenario, for example: not only bundles, but also operating system packages must be handled. With this purpose in mind, we have included the capability to delegate some deployment tasks to layer specific managers in Resolvit's architecture.

V*ersions of resources must be distinguished* for practical purposes. Software evolves over time and the version information will be important for defining dependencies with other elements. The resolution process must be able to read this information, either by a local check on the platform or querying an external repository. This requirement becomes even more important when the platform execution environment allows different versions of the same resource to coexist, as is the case of OSGi, for example.

Deployment engines *should be distributable*. Different reasons can force the distribution in the deployment processes: constraints in resources (the target platforms may be limited in resources and then it will be impossible for them to execute complex deployment tasks), security, business models and so on.

Context conditions are a source of valuable information for the deployment process. Therefore, the deployment engine *should be context-aware*. There are different sources of information that could be relevant for a context aware deployment: user preferences, the platform location, system features (architecture, network links, secondary storage size, CPU speed, etc.) or available resources (memory, disk space and so on.), just to mention some of them.

In the home services gateway scenario two other factors must be considered: security conditions and service costs. The execution of deployment tasks by unauthorized users might leave the platform unusable, hence the deployment engine must be executed under certain security constraints. Integration of cost models on the deployment engine is the other master key to deployment on service gateways. The feasibility of business models in this environment strongly depends on this. Ideally, the user should aid the deployment process with regards to cost preferences, as assumptions on this context are difficult to make, for example: not all users will prefer the cheapest option.

Deployment tasks *must be reliable*, changes carried out on a platform during deployment make a persistent change in its status. Certain errors might make the platform crash. For the sake of robustness, a deployment engine should either finish its tasks successfully or leave the platform in the same state it was before the deployment process was launched. In order to tackle this difficult problem we propose to divide the deployment process into two phases: check and execution. In the first one, all the information available (dependencies, user preferences, platform features, already deployed resources and so on) is analysed. As a result, different solutions to the deployment request are obtained. One of them will be selected according to certain policy and only then the second phase, execution, will be carried out atomically.

### 3.3. The service gateway scenario

In our reference scenario, the bundles will contain: service interfaces, their implementations and one deployment descriptor. The deployment descriptor will also be available separately in repositories, in order to avoid useless bundle downloads during the execution of the check phase.

The deployment process that we propose is launched with a first bundle as root, then recursively a search is carried out over all resources related by dependencies. This first step will create the set of possible deployment tasks to be carried out. We intend to cover all the deployment needs for service gateways, therefore the following deployment units have to be handled: bundles, native packages and device drivers. Once one of the possible configurations is selected the execution phase begins, according to certain customizable policies.

## 4. Resolvit's deployment metadata and resolution algorithm

In this section we are going to show how we have mapped the previously described requirements to create a deployment descriptor format for Resolvit, one of the key issues on deployment for SOA is dependency resolution. We will explain how Resolvit's dependency model is translated to the descriptor files and then we will explain the algorithm that it uses for deployment.

Dependencies must be explicitly and systematically described. By "systematically" we are referring to the fact that they must be available in a computer understandable format. This enables automated deployment engines to be implemented, avoiding manual tasks, which are costly in terms of required resources and in addition, prone to errors.

Resolvit's deployment metadata have been defined using an XML schema, its contents are classified as follows:

- Self describing information: deployment unit name, services offered, version, provider and the priority level.
- Dependencies with other entities: type of dependency, required service, cardinality, URL of the endpoint repository to find the required interface and so on.
- Information on required resources: disk space, processor architecture, operating system needed and so on.

### 4.1. Resolvit's dependency model

The dependency resolution process depends on the type of dependencies that we are capable of handling. Each type of dependency will somehow relate one source with one or more endpoints. With the aim of modelling dependencies generically, we have adopted a model that is based on Boole's algebra operators:

- AND, this type will be used to link a source with endpoint elements that are mandatory for the source's correct execution. Endpoints will have to be deployed before the source on the platform.
- NOT, whenever an identified conflict between a source and a certain endpoint is known. If the endpoint resource is already deployed on the platform, then the deployment engine will have to take a decision, because both elements cannot live together.
- OR, this type will be used when one or more resources can be selected at the same time.
- XOR, this type will be used when either the source or the endpoint may be chosen, but not at the same time.

One of the advantages of this dependency model is that is independent of the exact attributes that describe the sources or the endpoints of the dependencies. This allows the exact format of the deployment file to be updated if new context attributes are needed.

### 4.2. Resolvit's deployment algorithm

Resolvit's deployment algorithm is divided into two phases (see the activity diagram in Figure 2): dependencies resolution and execution respectively. The *first of the phases executes* the following steps *recursively*:

1. Read the deployment descriptor of the deployment unit and go to step 2.
2. Check the platform status, if the required resource is already installed the process finishes here. On the contrary, or if the resource is not up to date (see version information) go to step 3.
3. Read dependencies and go to the next step.
4. If there are no dependencies, go to step 5, else add them to the dependency tree - the data structure created to store the information collected on this phase- and go to step 6.
5. Iterate over each dependency relationship and search for endpoints deployment descriptors. If the dependency type is "NOT" then its information is included into the dependency tree, but it is not recursively evaluated.

In order to make the diagram easier to understand, only these steps are included, thus the next steps are only sketched:

6. In the dependency tree built during the previous phase, apply a selection policy and go to step 7. Step 5 will give the whole set of theoretically possible solutions as result, which will not be unique if optional relationships are detected.
7. Solve conflicts and install resources. A certain order has to be followed to guarantee the correct execution of all of them. This order begins with the chosen leaves of the tree that are at the bottom and climbs up the tree to the source of the deployment request.

## 5. Related Work

Deployment and configuration tasks are not new to the software community. Nonetheless, deployment is still nowadays an active field of research. Significant industrial and academic organizations are combining their efforts in adding new functionalities, proposing advanced models or enhancing the reliability of deployment activities. This section is organized as follows: first we talk about related deployment systems available in the industry, then about related research works and finally, about existing standards for component deployment.

Let us start with deployment systems already available in the industry. Installshield[INST 04] is one of the best examples of existing commercial tools targeted to application deployment. This type of tools are usually optimized for creating standalone installer archives and consequently do not fit well to dynamic environments.

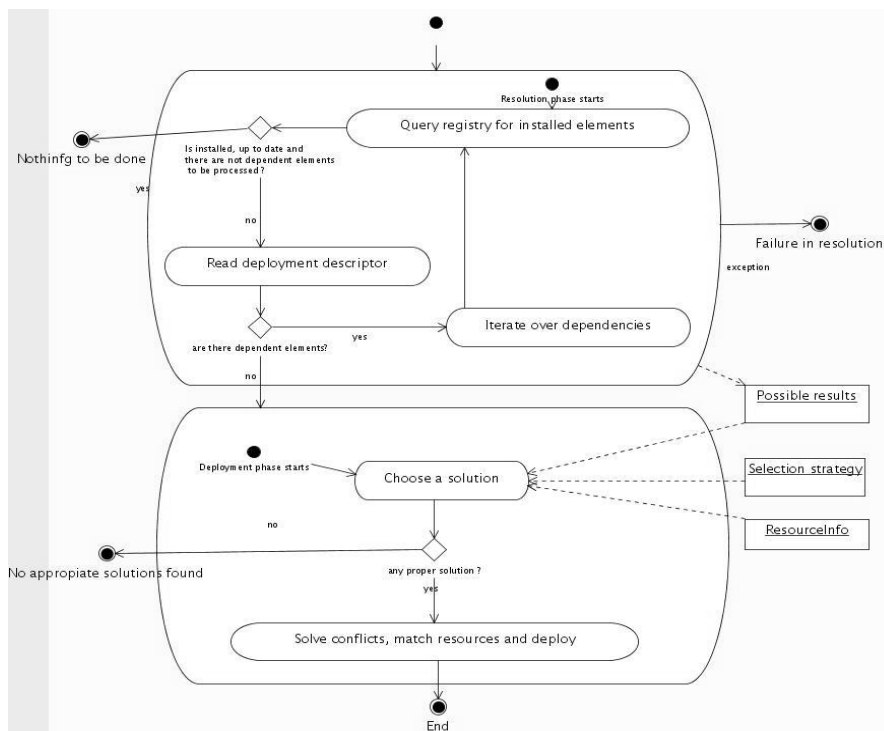

*Figure 2. Resolvit's deployment diagram*

Operating system vendors have traditionally been worried about maintaining a flexible and automatic infrastructure for configuration and deployment. MS Windows Update[MIC 04] can be mentioned, as one of the most widely spread commercial solutions. It allows MS Windows users to keep their operating systems up to date, by means of on line facilities. Software update packages are automatically downloaded and installed on the users' terminals. The release of .Net technologies has improved the scenario by adding the capability to maintain different versions of components and libraries running at the same time. .Net is a promising technology, but its evolution will probably be slowed by the requirement of compatibility with the previous MS Windows component models.

Debian is one of the first GNU/Linux distributions with tool support to cover most of the deployment tasks: dependencies resolution, installation, uninstallation, configuration and update of packages. In fact, the approach followed by Debian has been the source of inspiration for our work on deployment issues. Debian packages are stored in repositories, each package is described by a text file in terms of: self description (name, version, maintainer, size and so on), dependencies with other packages (required packages and known conflicts) and platform features (machine architecture, operating system kernel version, etc.). Currently, Debian is one of most stable and reliable solutions, as witnessed by the number of organizations that have chosen it as operating systems for their servers[DEB 04b]. This is not only because of the stability of the OS, but also for the capabilities to automate deployment activities, for example to apply security patches to a running web server. One of the valuable aspects of our work has been to broaden the approach of Debian, from deployment of Debian packages to deployment units in SOA.

Now let us talk about other significant related research activies such as ORYA[LES 03] or Software Dock[HAL 99], which try to cope with deployment of applications in distributed and heterogeneous environments. Both of them handle the whole deployment life cycle, by means of defining specialized deployment servers that carry out adaptation tasks and dependency resolution. Their conceptual approach is quite similar, but ORYA is more flexible than Software Dock in enforcing different deployment strategies. This flexibility derives from the composable deployment model followed by ORYA. Deployment processes are specified by an ordered set of basic activities and/or other deployment processes. ORYA has defined a dependencies model built in terms of required elements and software conflicts, the latter are expressed as software constrains. Our dependency model covers these categories with the AND and NOT dependency types. ORYA´s dependency model lacks the expressiveness to handle optional elements, which is considered in our model through the XOR and OR dependency types.

Finally, let us talk about deployment standards for component-based applications. Deployment and configuration for distributed component-based applications is addressed by an OMG standard[OMG 03]. The body of the standard is divided into two parts, the first one presents a platform independent model (PIM) and the second one, includes the transformation from the PIM to a platform specific model (PSM), for the CORBA component model (CCM).

The PIM is divided into three levels: component level, target level and execution level. The entities that appear at each of the levels have been classified into two groups: data entities and process entities. Both groups are closely related, the data entities are used to model the input and output parameters for the methods of the process entities. Process entities are used to model deployment business logic tasks. A detailed explanation of this standard is out of the scope for this article, but the following remarks are worth mentioning for their significance in the deployment field:

– A deployment algorithm has been proposed as a recommendation.
– It includes and defines the set of roles involved in the deployment process. The roles are classified with regard to the three levels considered in the standard: component, target and execution.
– Nodes and connections are explicitly part of the model. This allows deployment plans to be defined, taking into account the network topology.
– Adaptive deployment is a goal for the specification. Node and link resources can be formally described and considered in a deployment plan.
– It includes a catalogue of exceptions that compiles errors that might happen during the deployment process. However, no strategies for handling these exceptions are proposed in the standard. One error during some deployment tasks may lead to an unstable platform. Ideally, deployment tasks should be executed as transactions, but this is something quite difficult to get. ORYA tries to work around this problem coupling each "do" deployment task with an "undo" task.

## 6. Conclusions and future work

The main contributions of this article are as follows: we have provided a general overview on the field of deployment for dynamic environments. In this context, an analysis on the main goals for a deployment engine has been provided. We have also introduced some parts of Resolvit. Resolvit is a deployment engine targeted to solve the needs of OSGi enabled service gateways. Its dependency model and the format and contents of the deployment descriptors it uses have been explained to illustrate how the requirements on deployment have been mapped onto a specific and industrially realistic scenario.

Although ORYA, Software Dock and the OMG standard use different models, all of them share a similar flavour. Deployment is tackled as a top down process, where certain entities create a plan, making use of specialized servers. The execution of deployment is carried out from these controlling actors to passive target platforms. Our approach is quite different. Somehow it is related to applications that follow the p2p (peer to peer) architecture. We conceive deployment as a flexible and automated task that has to be carried out with the focus on the deployment target. In fact, Resolvit was initially thought to be a kind of intelligent agent for a service platform, that with the knowledge of the platform status dynamically looks up and deploys applications.

The scope of our work, service deployment for home service gateways has influenced on our approach. Other existing deployment engines have focused on solving the deployment requirements for big companies, while our interest so far has been: end users homes. In this context, the final user needs a cost aware deployment mechanism. This requirement has been taken into account on the definition of Resolvit's architecture.

The approach presented here is being validated in an end-to-end services demonstrator carried out within the EUREKA ITEA Osmose[OSM 04] project. This project is currently running and we expect to have better validation results at the end of it. There are some issues that we still have in mind to improve the deployment engine architecture, such as adding an intelligent metadata repository, instead of specifying the exact URL in the descriptors or implementing and validating new selection and conflict resolution policies.

**7. Bibliography**

[BIE 04] G. Bieber and J. Carpenter. *Introduction to Service-Oriented Programming*, September 2001, http://www.openwings.org/download.html

[BOU 03] C. Boulton. *Service-Oriented Architectures Underpin On-demand*, 26th May 2003, Digital Magazine: InternetNews.com: http://www.internetnews.com/ent-news/article.php/2212131.

[CER 03] H.Cervantes and R.S. Hall. *Automating Service Dependency Management in a Service-Oriented Component Model*, in the proceedings of ICSE CBSE6 Workshop 2003, held in Portland, USA.

[DEB 04a] Debian, official web site at: http://www.debian.org

[DEB 04b] List of organizations using Debian, http://www.debian.org/users

[ECL 04] Eclipse, official home page: http://www.eclipse.org

[HAL 99] R. S. Hall, Heimbigner and A. L. Wolf. *A cooperative approach to support software deployment using the software dock*, in the proceedings of ICSE, held at Los Angeles 1999. Published IEEE Computer Society Press, pages 174-183.

[INST 04] Installshield, home page: http://www.installshield.com/

[LES 03] V. Lestideau and N. Belkhatir. *Providing highly automated and generic means for software deployment Process*. European Workshop on Software Process Technology, Helsinki, Finland, September 1-2, 2003

[MIC 04] Microsoft Windows Update, http://windowsupdate.microsoft.com

[OMG 03] *Deployment and Configuration of Component-based Distributed Applications Specification*. OMG Draft Adopted Specification ptc/03-07-02. June 2003.

[OSG 03] Open Services Gateway Initiative. *OSGi Service Platform*, specification release 3.0, March 2003.

[OSM 04] EUREKA ITEA-Osmose, official web site at: http://itea-osmose.org